\documentclass[aps,prd,twocolumn,10pt,groupedaddress]{revtex4-1}
\usepackage{amssymb}
\usepackage{graphicx}
\usepackage{amsmath}
\usepackage{hyperref}
\usepackage{float}

\begin{document}

\title{Sensitivity of primordial black hole abundance on the reheating phase}
\author{Rong-Gen Cai}
\email{cairg@itp.ac.cn}
\author{Tong-Bo Liu}
\email{liutongbo@itp.ac.cn}
\author{Shao-Jiang Wang}
\email{schwang@itp.ac.cn}
\affiliation{CAS Key Laboratory of Theoretical Physics, Institute of Theoretical Physics, Chinese Academy of Sciences, Beijing 100190, China}
\affiliation{School of Physical Sciences, University of Chinese Academy of Sciences, No.19A Yuquan Road, Beijing 100049, P.R. China}
\date{\today}

\begin{abstract}
We investigate the sensitivity of reheating history on the abundance of primordial black holes (PBHs). Contrary to the monochromatic case of mass fraction of PBH, reheating era with different e-folding and equation-of-state can have a substantial impact on the abundance of PBH with an extended mass fraction. We demonstrate explicitly this reheating sensitivity in an illustrative model of single field inflation with a quasi-inflection point, and find that both the peak position and amplitude of the extended mass fraction as well as the abundance of PBH in DM can vary by many orders of magnitude, which adds another layer of uncertainty on the PBH scenarios as dark matter.
\end{abstract}
\maketitle

\section{Introduction}

All the possible indications for the very existence of dark matter (DM) come from the observations of astrophysics and cosmology, including the galactic scales (galaxy rotation curves \cite{Rubin:1970zza,Freeman:1970mx}), galaxy cluster scales (velocity dispersions \cite{Zwicky:1933gu,2017arXiv171101693A}, X-ray radiation, bullet cluster \cite{Squires:1995ee,Clowe:2006eq}) and cosmological scales (BBN: big bang nucleosynthesis \cite{Burles:2000zk}, CMB: cosmic microwave background \cite{Ade:2015xua}, N-body simulations \cite{Springel:2005mi}). It is worth noting that the modified Newtonian dynamics (MOND) can be unnecessary since one can even reproduce the MOND-like behavior by EAGLE's simulation in Lambda-cold-dark-matter ($\Lambda$CDM) model \cite{Dai:2017unr}. However, there is currently no appealing evidence for DM of particle nature \cite{Bertone:2004pz}, including the direct detections (LUX \cite{Akerib:2016vxi}, PandaX \cite{Cui:2017nnn}, XENON1T \cite{Aprile:2017iyp}, CDMS \cite{Agnese:2013rvf}, ADMX \cite{Du:2018uak}), indirect detections (ICECUBE \cite{Aartsen:2016zhm}, Pamela \cite{Adriani:2013uda}, AMS \cite{Aguilar:2016kjl}, DAMPE \cite{Ambrosi:2017wek}) and collider search (LHC). The above dilemma naturally leads us to an alternative possibility that DM might be of non-particle nature \cite{Axelrod:2016nkp}, and only participates the gravitational interaction. The most promising candidate along this direction is the primordial black holes (PBHs) \cite{Hawking:1971ei,Carr:1974nx,Carr:1975qj}. 

PBHs make a perfect candidate for DM since they are known as cold, stable and collisionless massive astrophysical compact halo object (MACHO), which could be generated deep inside the radiation-dominated era therefore free from BBN constraint on baryonic density. The abundance of PBH in DM has been constrained by means of the electromagnetic wave (EMW) over the past few decades \cite{Carr:2009jm}. See \cite{Carr:2016drx} for the monochromatic mass fractions and \cite{Carr:2017jsz} for the extended mass fractions. Different from the EMW, the gravitational wave (GW) from PBHs provides another powerful detections. Soon after the first GW detection GW150914 \cite{Abbott:2016blz}, there is a renewing and growing interest on PBH as DM by explaining the LIGO detection from PBHs merger \cite{Kashlinsky:2016sdv,Bird:2016dcv,Clesse:2016vqa,Sasaki:2016jop}. See for the most recent and comprehensive review on PBH \cite{Sasaki:2018dmp}.

The major concern of PBH scenario of DM is its uncertainty from the generation models, formation models, accretion models and merger models, where a lot of nasty physics can be involved to make the prediction less traceable and varied in general by orders of magnitude, for example, the shape of the peak \cite{Germani:2018jgr}, the choice of the window function \cite{Ando:2018qdb}, the effect from quantum diffusion \cite{Pattison:2017mbe,Biagetti:2018pjj,Ezquiaga:2018gbw}, to name a few. In this paper, we investigate the effect from reheating history on PBH (See \cite{Martin:2014nya,Dai:2014jja,Creminelli:2014fca,Munoz:2014eqa,Gong:2015qha,Cai:2015soa,Cook:2015vqa} for the effect from reheating history on inflation). Remarkably the abundance of PBH with monochromatic mass fraction is immune from the reheating history \cite{Pi:2017gih} . However, for extended mass fraction of PBH generated from the collapse of the re-entered primordial fluctuations, the sensitivity of reheating can have a substantial effect on the abundance of PBH as we will show explicitly in an illustrative model with a quasi-inflection point \cite{Garcia-Bellido:2017mdw} (See also \cite{Motohashi:2017kbs,Germani:2017bcs} for the violation of slow-roall approximation that was initially used in the first version of \cite{Garcia-Bellido:2017mdw}).

The paper is organized as follows: In Sec.\ref{sec:inflation}, the general formalism of primordial fluctuation is presented along with an illustrative model with quasi-inflection point. In Sec.\ref{sec:PBH}, the abundance of PBH with extended mass fraction is presented for different reheating histories. Section \ref{sec:conclusion} is devoted to conclusions.

\section{Single field inflationary model}\label{sec:inflation}

\subsection{An illustrative model}

\begin{figure*}
\centering
\includegraphics[width=0.49\textwidth]{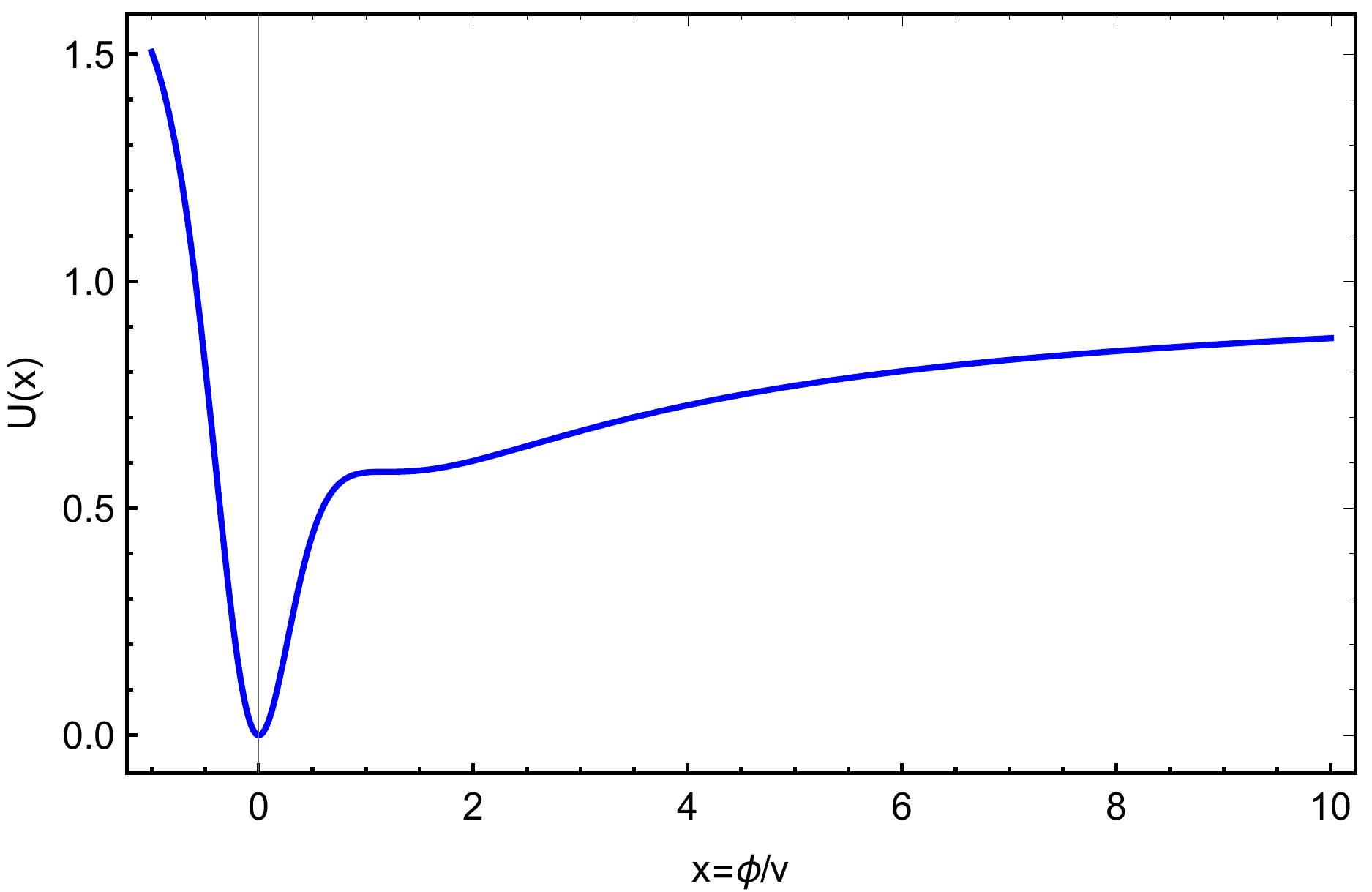}
\includegraphics[width=0.49\textwidth]{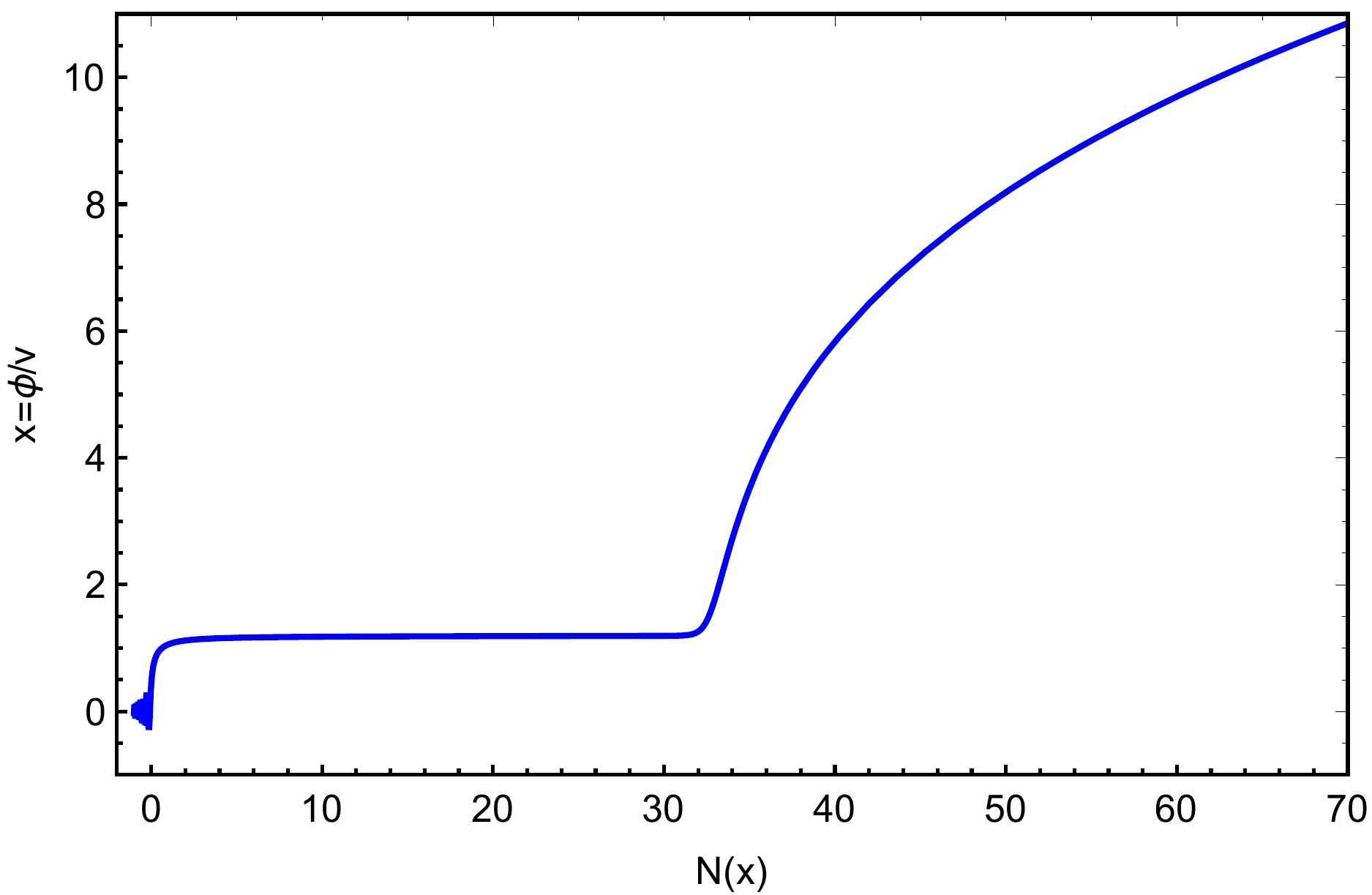}\\
\includegraphics[width=0.49\textwidth]{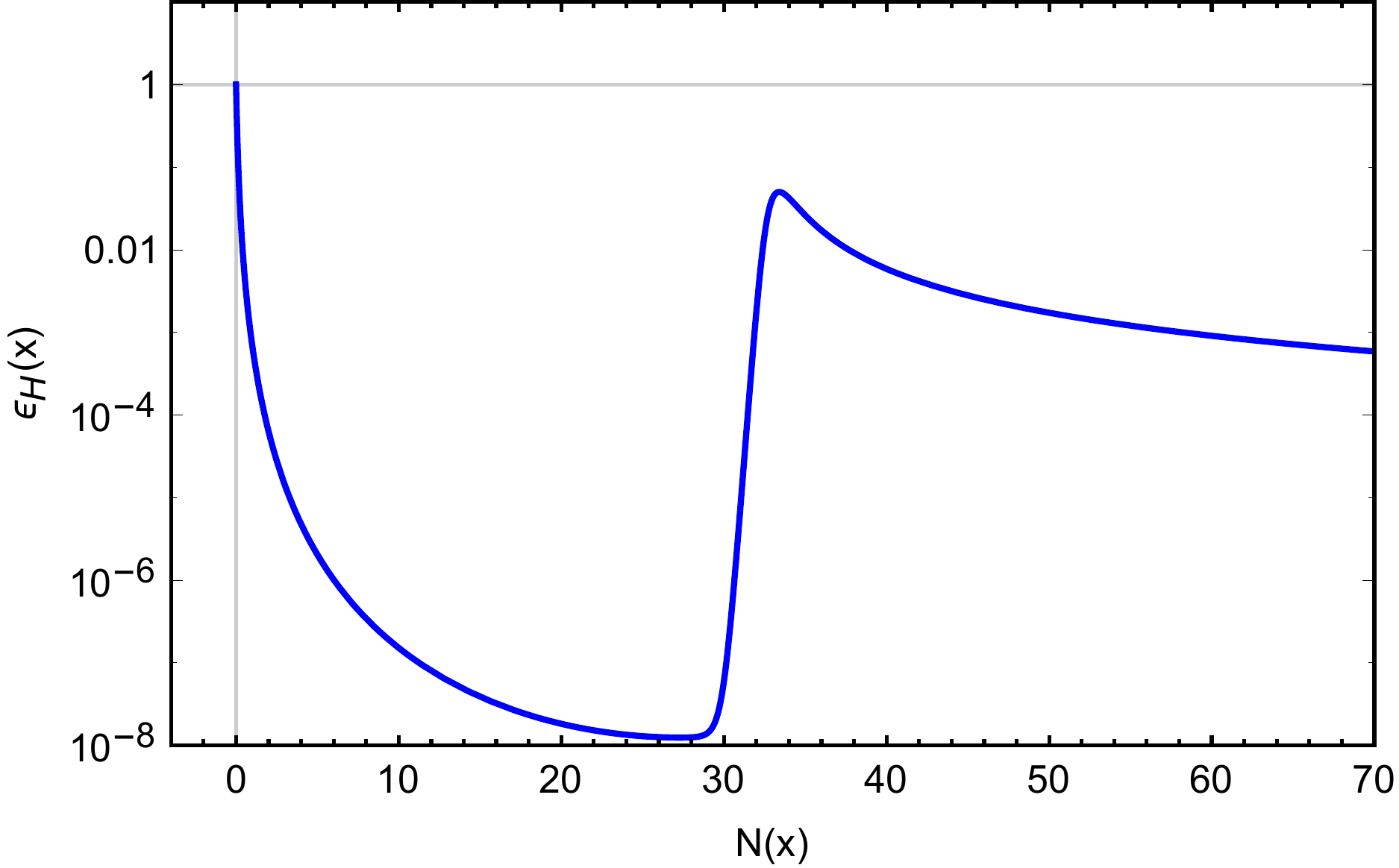}
\includegraphics[width=0.49\textwidth]{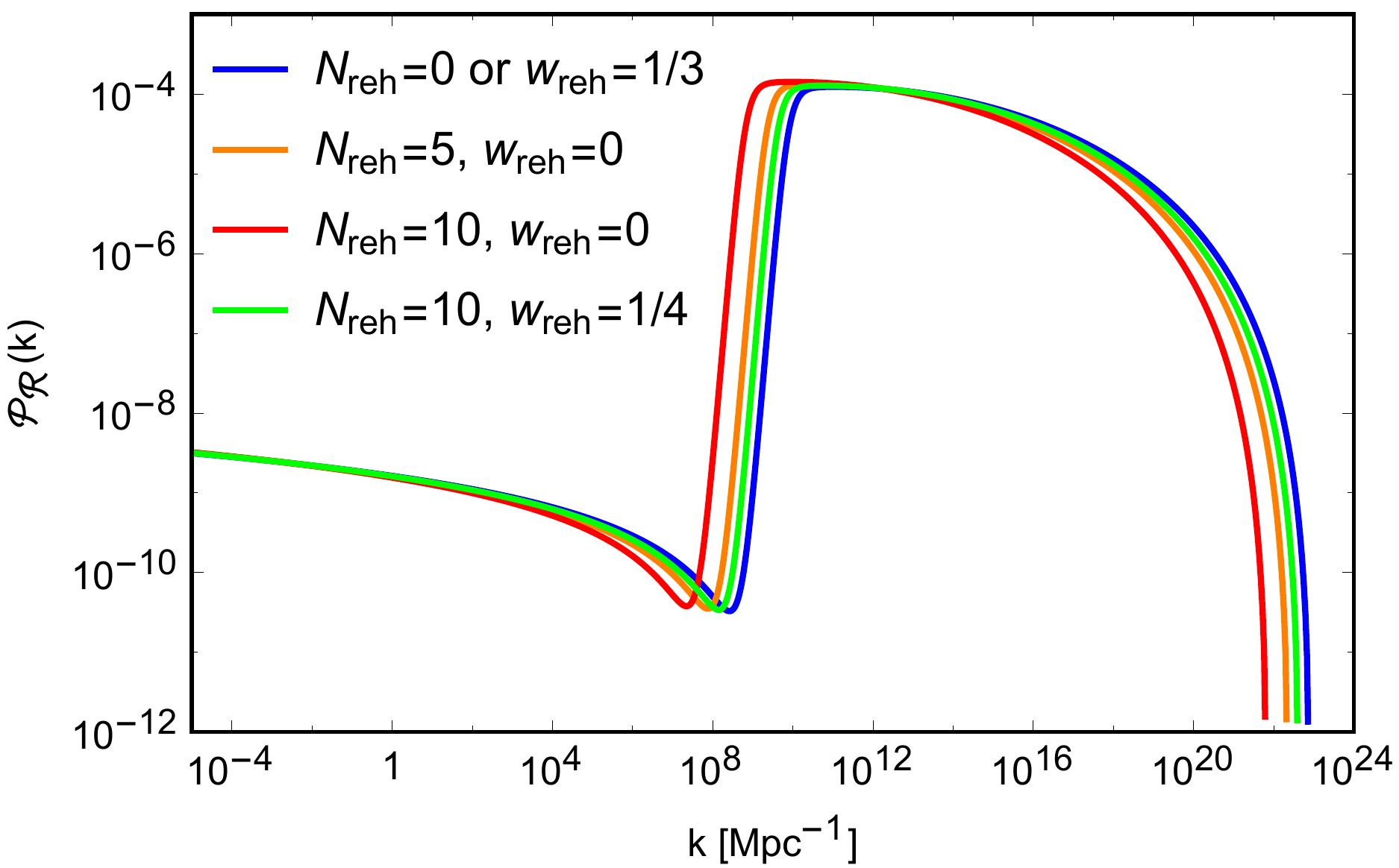}\\
\caption{Upper left panel: the asymptotic normalized and dimensionless part of the effective pontential with a quasi-inflection point. Upper right panel: the exact solution of inflationary dynamics with respect to the e-folding number. Lower left panel: the Hubble slow-roll parameter that saturates unity exactly at $N=0$. Lower right panel: the power spectrum the curvature perturbation with different reheating histories. }\label{fig:inflation}
\end{figure*}

In this section, we will take an illustrative model of inflation with a quasi-inflection point as introduced in \cite{Garcia-Bellido:2017mdw}. The effective potential is
\begin{align}
V(\phi)=\left(\frac12m^2\phi^2-\frac13\alpha v\phi^3+\frac14\lambda\phi^4\right)\left(1+\xi\phi^2\right)^{-2},
\end{align}
which, under the constraint $m^2=\lambda v^2$, the redefinition of variable $x=\phi/v$ and parameters $a=\alpha/\lambda, b=\xi v^2$, can be recast in dimensionless form
\begin{align}
V(x)=\frac{\lambda v^4}{4b^2}U(x)=\frac{\lambda v^4}{12}\frac{x^2(6-4ax+3x^2)}{(1+bx^2)^2},
\end{align}
where $\lambda v^4/4b^2$ is the asymptotic value of $V(x)$ so that
\begin{align}
U(x)=\frac{b^2x^2(6-4ax+3x^2)}{3(1+bx^2)^2},
\end{align}
is normalized as 1 in large $x$ limit. When the parameter $b$ acquires a critical value
\begin{align}
b_c&=1-\frac{a^2}{3}+\Delta(a),\\
\Delta(a)&=\frac{a^2}{3}\left(\frac{9}{2a^2}-1\right)^\frac23,
\end{align}
the effective potential admits an inflection point at
\begin{align}
x_0&=\frac{b_c-1+\sqrt{(b_c-1)^2+a^2b_c}}{ab_c}.
\end{align}
To have a quasi-inflection point, one should shift the parameter $b=b_c-\beta$ by a small amount $\beta$ from its critical value $b_c$. In what follows, we will take the same fiducial values $a=1, \beta=10^{-4}, \kappa^2v^2=0.108$ as in \cite{Garcia-Bellido:2017mdw}, of which the dimensionless part of effective potential is presented in the upper left panel in Fig.\ref{fig:inflation}.

\subsection{Equation-of-motion}

To be general, we take an inflation potential $V(\phi(t);p)$ with a free parameter $p$ to work out the power spectrum. The equation-of-motion (EOM) is
\begin{eqnarray}
\ddot{\phi}&+&3H\dot{\phi}+V'(\phi;p)=0;\label{eq:inflaton}\\
H^2&=&\frac{\kappa^2}{3}\left(\frac12\dot{\phi}^2+V(\phi;p)\right);\label{eq:FLRW1}\\
\dot{H}&=&-\frac{\kappa^2}{2}\dot{\phi}^2.\label{eq:FLRW2}
\end{eqnarray}
Here the reduced Planck mass $1/\kappa=2.435\times10^{18}\,\mathrm{GeV}\equiv M_\mathrm{Pl}$. Using $dN=-Hdt$, and noting that
\begin{eqnarray}
\dot{\phi}&=&-H\phi'(N);\label{eq:phidot}\\
\ddot{\phi}&=&-\dot{H}\phi'(N)+H^2\phi''(N);\\
           &=&\frac{\kappa^2}{2}\dot{\phi}^2\phi'(N)+H^2\phi''(N);\\
           &=&\frac{\kappa^2}{2}H^2\phi'(N)^3+H^2\phi''(N),
\end{eqnarray}
the EOM becomes
\begin{align}
\phi''(N)+\frac{\kappa^2}{2}\phi'(N)^3-3\phi'(N)+\frac{V'(\phi;p)}{H^2}=0
\end{align}
Note that the Hubble parameter can be written in the form of
\begin{equation}\label{eq:Hubble}
H(N;p)^2=\frac{\frac{\kappa^2}{3}V(\phi(N);p)}{1-\frac{\kappa^2}{6}\phi'(N)^2}.
\end{equation}
with the help of \eqref{eq:FLRW1} and \eqref{eq:phidot}. For some inflation potential, the combination $V'(\phi;p)/V(\phi;p)$ could be independent of the free parameter $p$, namely $p$ is an overall factor in the potential, then the EOM is
\begin{equation}
\phi''(N)+\frac{\kappa^2}{2}\phi'(N)^3-3\phi'(N)
+\frac{1-\frac{\kappa^2}{6}\phi'(N)^2}{\frac{\kappa^2}{3}}\frac{V'}{V}(\phi)=0
\end{equation}
In our illustrative model, one can choose the free parameter to be $\lambda$, and fix all other parameters in the effective potential at their fiducial values. Now the dimensionless EOM is
\begin{align}
x''(N)+\frac{(\kappa v)^2}{2}x'(N)^3-3x'(N)+\frac{1-\frac{(\kappa v)^2}{6}\phi'^2}{\frac{(\kappa v)^2}{3}U(x)}\frac{\mathrm{d}U(x)}{\mathrm{d}x}=0
\end{align}
Even though the inflationary dynamics admits attractor, solving above EOM from an arbitrarily chosen initial condition
\begin{equation}
\phi(N_i)=\phi_i, \quad \phi'(N_i)=\phi'_i
\end{equation}
with sufficiently large $N_i$ does not necessarily end the inflation exactly at $N=0$, namely the the Hubble slow-roll parameter
\begin{equation}
\epsilon_H(N)=-\frac{\dot{H}}{H^2}=\frac{\kappa^2}{2H^2}\dot{\phi}^2=\frac{\kappa^2}{2}\phi'(N)^2,
\end{equation}
can be still smaller than unity at $N=0$ with solved solution $\phi(N)$. One can either shift the solution $\phi(N+N_f)$ with amount $N_f$ obtained from $\epsilon_H(-N_f)=1$, or carefully increase the initial velocity $\phi'_i$ until $\epsilon_H(0)=1$. In our illustrative model, we choose the initial conditions $N_i=80, x_i=12, x'_i=0.65110936$ so that the Hubble slow-roll parameter saturates unity exactly at $N=0$ as shown in the lower left panel of Fig.\ref{fig:inflation}. The obtained exact solution $x(N)$ is presented in the upper right panel of Fig.\ref{fig:inflation}. 

\subsection{Power spectrum}

To fix the pivot scale of primordial fluctuations that re-enters the Hubble horizon at last scattering surface, we use the WMAP normalization conditions \cite{Hinshaw:2012aka}
\begin{align}
&k_\mathrm{CMB}=a_\mathrm{CMB}H_\mathrm{CMB}=0.002\,\mathrm{Mpc^{-1}};\label{eq:CMB1}\\
&A_s=\mathcal{P}_\mathcal{R}|_{N=N_\mathrm{CMB}}=2.4\times10^{-10}.\label{eq:CMB2}
\end{align}

The first CMB normalization condition can be evaluated with respect to the current Hubble scale, 
\begin{align}
\frac{k_\mathrm{CMB}}{a_0H_0}=\frac{a_\mathrm{CMB}}{a_\mathrm{end}}\frac{a_\mathrm{end}}{a_\mathrm{reh}}\frac{a_\mathrm{reh}}{a_0}\frac{H_\mathrm{CMB}}{H_0}.
\end{align}
Here we adopt the fiducial value $h=0.7$ for Hubble parameter, and the unit conversion $1\,\mathrm{GeV}=2.488767\times10^{37}\times2\pi\,\mathrm{Mpc}^{-1}, 1\,\mathrm{Mpc}^{-1}=4.018054*10^{-38}/2\pi\,\mathrm{GeV}$. Therefore $H_0=1.49215\times10^{-42}\,\mathrm{GeV}=0.000233\,\mathrm{Mpc}^{-1}$.

The duration between the exit of pivot scale and the end of inflation is denoted as $N_\mathrm{CMB}$, and the duration between the end of inflation and the end of reheating is denoted as $N_\mathrm{reh}$, then
\begin{eqnarray}
\frac{a_\mathrm{CMB}}{a_\mathrm{end}}&=&\mathrm{e}^{-N_\mathrm{CMB}};\\
\frac{a_\mathrm{end}}{a_\mathrm{reh}}&=&\mathrm{e}^{-N_\mathrm{reh}};\\
\frac{a_\mathrm{reh}}{a_0}&=&\left(\frac{43}{11g_{\mathrm{reh}}}\right)^{\frac{1}{3}}\frac{T_{\gamma}}{T_{\mathrm{reh}}},\label{eq:Treh}
\end{eqnarray}
where in third line we use the conservation equation of entropy $g_{\mathrm{reh}}a_{\mathrm{reh}}^3T_{\mathrm{reh}}^3=g_{\gamma}a_0^3T_{\gamma}^3+g_{\nu}a_0^3T_{\nu}^3=(43/11)T_{\gamma}^3a_0^3$ by noting that
$g_{\gamma}=2$, $g_{\nu}=(7/8)\times3\times2=21/4$, and $T_{\nu}^3=(4/11)T_{\gamma}^3$ with $T_{\gamma}=2.7255\,\mathrm{K}=2.7255\times8.61733\times10^{-14}\,\mathrm{GeV}$. The number of degrees of freedom is usually taken as $g_\mathrm{reh}=106.75$.

To compute the reheating temperature in \eqref{eq:Treh}, one assumes that the whole history during the reheating phase can be effectively described by the $e$-folding number $N_{\mathrm{reh}}$ and the effective equation-of-state (EoS) parameter $w_{\mathrm{reh}}$ \cite{Cai:2015soa}, thus we are able to relate the reheating phase to the inflationary phase via
\begin{eqnarray}
\rho_{\mathrm{reh}}&=&\frac{\pi^2}{30}g_{\mathrm{reh}}T_{\mathrm{reh}}^4;\\
\rho_{\mathrm{reh}}&=&\rho_{\mathrm{end}}\,\mathrm{e}^{-3N_{\mathrm{reh}}(1+w_{\mathrm{reh}})};\\
\rho_{\mathrm{end}}&=&\frac{3}{\kappa^2}H(0;p)^2
\end{eqnarray}
where the end of inflation is identified as $\phi(N=0)$ obviously. Combining the above three equations gives rise to
\begin{equation}
T_{\mathrm{reh}}=\left(\frac{30}{\pi^2g_\mathrm{reh}}\right)^\frac14
\left(\frac{3}{\kappa^2}H(0;p)^2\mathrm{e}^{-3N_{\mathrm{reh}}(1+w_{\mathrm{reh}})}\right)^\frac14.
\end{equation}

Therefore the first CMB normalization condition \eqref{eq:CMB1} can be expressed as
\begin{align}\label{eq:CMB1new}
k(N_\mathrm{CMB},p)=\frac{\mathrm{e}^{-N_\mathrm{CMB}-N_\mathrm{reh}}\left(\frac{43}{11g_{\mathrm{reh}}}\right)^{\frac{1}{3}}T_{\gamma}H(N_\mathrm{CMB};p)}{\left(\frac{30}{\pi^2g_\mathrm{reh}}\right)^\frac14
	\left(\frac{3}{\kappa^2}H(0;p)^2\mathrm{e}^{-3N_{\mathrm{reh}}(1+w_{\mathrm{reh}})}\right)^\frac14}
\end{align}
for given solution $\phi(N)$, unknown quantities $N_\mathrm{CMB}$ and $p$ as well as the reheating history  $N_\mathrm{reh}$ and $w_\mathrm{reh}$. The second CMB normalization condition can be evaluated as
\begin{align}\label{eq:CMB2new}
A_s(N_\mathrm{CMB},p)=\frac{1}{4\pi^2}\frac{H(\phi(N_\mathrm{CMB});p)^2}{\phi'(N_\mathrm{CMB})^2},
\end{align}
for given solution $\phi(N)$ and unknown quantities $N_\mathrm{CMB}$ and $p$.

Now, directly solving the CMB normalization conditions \eqref{eq:CMB1new} and \eqref{eq:CMB2new} gives rise to the e-folding number $N_\mathrm{CMB}$ when pivot scale exits horizon and the free parameter $\lambda$ that renders observed amplitude of curvature perturbation. In our illustrative model, we find $N_\mathrm{CMB}=59.12, \lambda=4.39\times10^{-7}$ for $N_\mathrm{reh}=0$ or $w_\mathrm{reh}=1/3$, and $N_\mathrm{CMB}=57.88, \lambda=4.71\times10^{-7}$ for $N_\mathrm{reh}=5, w_\mathrm{reh}=0$, and $N_\mathrm{CMB}=56.65, \lambda=5.06\times10^{-7}$ for $N_\mathrm{reh}=10, w_\mathrm{reh}=0$, and $N_\mathrm{CMB}=58.50, \lambda=4.54\times10^{-7}$ for $N_\mathrm{reh}=10, w_\mathrm{reh}=1/4$. The corresponding power spectrum of curvature perturbation is presented in the lower right panel of Fig.\ref{fig:inflation}, where the peak position at small scales is shifted to larger scale for increasing $N_\mathrm{reh}$ and fixed $w_\mathrm{reh}$, and is shifted to smaller scale for increasing $w_\mathrm{reh}$ and fixed $N_\mathrm{reh}$.

\section{Primordial black hole production}\label{sec:PBH}

The power spectrum of curvature perturbation of our illustrative model manifests a broad peak at small scales, therefore the corresponding PBHs collapsed from these primordial fluctuations at re-entry admit extended mass fraction.

When sufficiently large density fluctuations at a certain scale enters the Hubble horizon, PBH is thus formed as result of the sufficient overdense region. The mass of PBH collapsed from primordial fluctuations at re-entry is usually approximated by the horizon mass at formation,
\begin{align}\label{eq:MPBH}
M_\mathrm{PBH}^\mathrm{form}=\gamma\frac43\pi\rho_\mathrm{crit}^\mathrm{form}H_\mathrm{form}^{-3}
=\gamma\frac{4\pi M_\mathrm{Pl}^2}{H_\mathrm{form}},
\end{align}
where the correction factor $\gamma\simeq0.2$ \cite{Carr:1975qj}. It will be shown in \ref{subsec:monochromatic} that the mass of PBH at formation is independent from reheating history for the monochromatic case, so is its abundance. However, the abundance of PBH with extended mass fraction depends crucially on different reheating histories, due to the dramatical change of the relative abundance of each monochromatic component of PBHs. The mass fraction is defined below.

The mass fraction of PBH $\beta^\mathrm{form}(M)$ is defined as the fraction of Universe collapsed into PBH of mass $M$ at formation, which can be regarded as the probability distribution function $P(\delta)$ integrated over the density contrast $\delta$ that is larger than a certain threshold $\delta_c$. The probability distribution function of density perturbations is usually assumed to be Gaussian, therefore the mass fraction of PBH can be evaluated as
\begin{align}
\beta^\mathrm{form}(M)=\int_{\delta_c}^\infty\frac{\mathrm{d}\delta}{\sqrt{2\pi}\sigma}\mathrm{e}^{-\frac{\delta^2}{2\sigma^2}}=\frac12\mathrm{erfc}\left(\frac{\delta_c}{\sqrt{2}\sigma}\right).
\end{align}
The variance of density perturbations $\sigma$ is given by the primordial power spectrum of curvature perturbations via $\sigma^2=\langle\delta^2\rangle=\mathcal{P}_\mathcal{R}(k_{N_\mathrm{PBH}})$ evaluated at the scale $k_{N_\mathrm{PBH}}$ that re-enters the Hubble horizon at the formation of PBH of mass $M(N_\mathrm{PBH})$. The variance of density perturbations is certainly much smaller than the threshold of PBH formation, therefore the mass fraction of PBH is further approximated as
\begin{align}
\beta^\mathrm{form}(M)\approx\frac{\sigma}{\sqrt{2\pi}\delta_c}\mathrm{e}^{-\frac{\delta_c^2}{2\sigma^2}}.
\end{align}

\subsection{The monochromatic case}\label{subsec:monochromatic}

\begin{figure}
\centering
\includegraphics[width=0.5\textwidth]{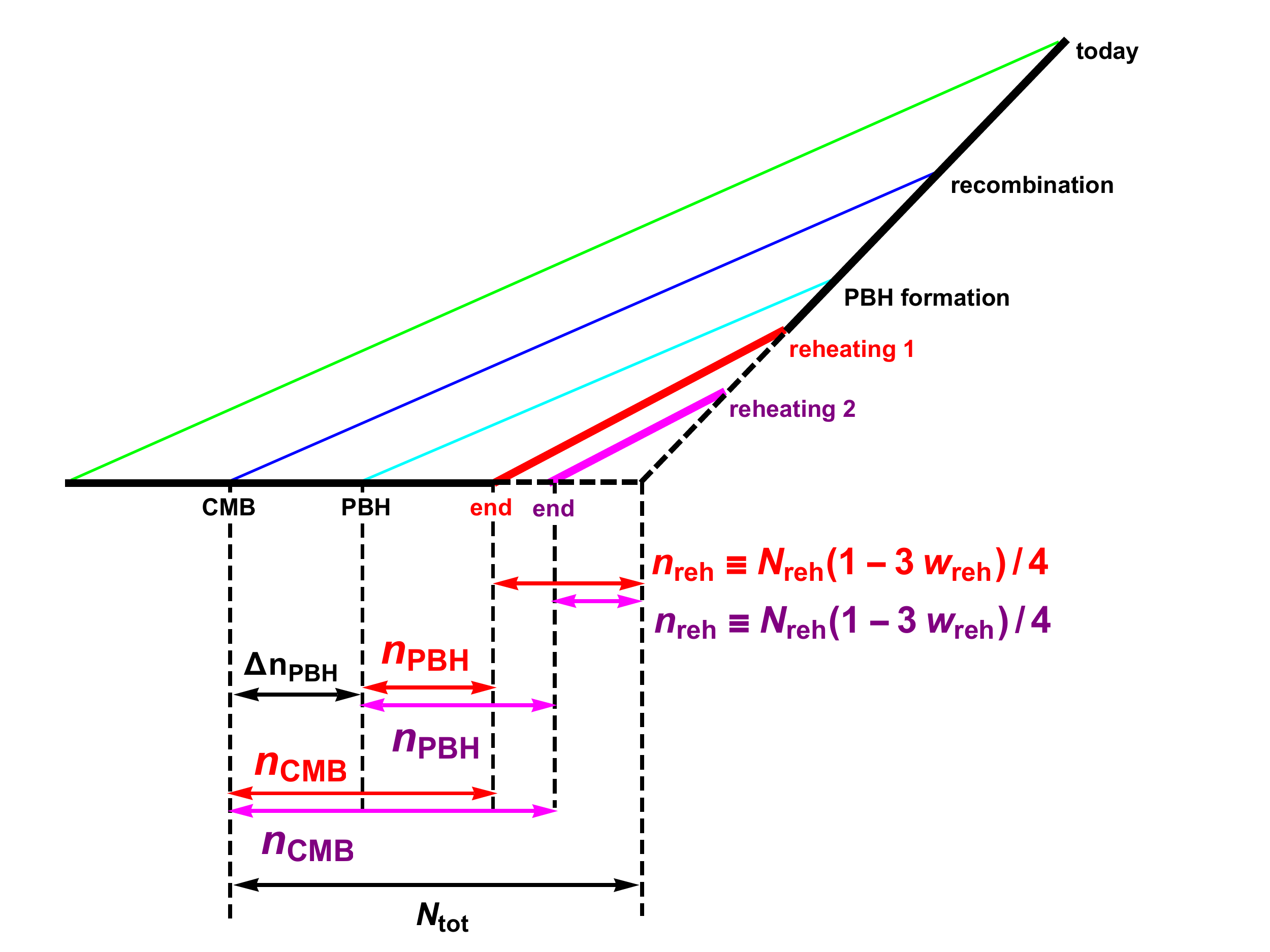}\\
\caption{Schematic illustration on the insensitivity of reheating history for monochromatic PBH. The PBH mass depends only on the $N_\mathrm{tot}-\Delta n_\mathrm{PBH}$, which is independent of reheating histories presented as red and purple lines, even though both $N_\mathrm{CMB}$ and $N_\mathrm{PBH}$ rely on the reheating histories. It is worth noting that the end of inflation is fixed at vanishing e-folding $N=0$.}\label{fig:monochromatic}
\end{figure}

For monochromatic case, the mass of PBH at formation is independent from reheating history. To see this, one use the scaling relation $H\propto a^{-\frac32(1+w)}$ to relate the scale factor at PBH formation with the scale factor at exit of primordial fluctuation with e-folding number $N_\mathrm{PBH}$,
\begin{align}\label{eq:aform}
\frac{a_\mathrm{form}}{a(N_\mathrm{PBH})}
=\mathrm{e}^{2\left[N_\mathrm{PBH}+\frac14N_\mathrm{reh}(1-3w_\mathrm{reh})\right]}
\frac{H_\mathrm{end}}{H(N_\mathrm{PBH})}.
\end{align}
Therefore the Hubble parameter at formation of PBH can be related to the Hubble parameter at exit of primordial fluctuation \cite{GarciaBellido:1996qt} via $a_\mathrm{form}H_\mathrm{form}=a(N_\mathrm{PBH})H(N_\mathrm{PBH})$, namely
\begin{align}\label{eq:Hform}
\frac{H_\mathrm{form}}{H(N_\mathrm{PBH})}
=\mathrm{e}^{-2\left[N_\mathrm{PBH}+\frac14N_\mathrm{reh}(1-3w_\mathrm{reh})-\frac12\ln\frac{H(N_\mathrm{PBH})}{H(N=0)}\right]}.
\end{align} 
It is worth noting that, the combination $\frac14N_\mathrm{reh}(1-3w_\mathrm{reh})$ in the exponential factor in \eqref{eq:Hform} can be expressed as
\begin{align}
n_\mathrm{reh}
&\equiv N_\mathrm{reh}-\frac12\ln\frac{H(N_\mathrm{reh})}{H(N=0)};\\
&=N_\mathrm{reh}-\frac12\int^{N_\mathrm{reh}}_0\epsilon_H(N)\mathrm{d}N;\\
&=N_\mathrm{reh}-\frac12\int^{N_\mathrm{reh}}_0\frac32(1+w_\mathrm{reh})\mathrm{d}N;\\
&=\frac14N_\mathrm{reh}(1-3w_\mathrm{reh}).
\end{align}
Similarly after introducing the notations
\begin{align}
n_\mathrm{CMB}&\equiv N_\mathrm{CMB}-\frac12\ln\frac{H(N_\mathrm{CMB})}{H(N=0)};\\
n_\mathrm{PBH}&\equiv N_\mathrm{PBH}-\frac12\ln\frac{H(N_\mathrm{PBH})}{H(N=0)},
\end{align}
the exponential factor in \eqref{eq:Hform} can be expressed as
\begin{align}
N_\mathrm{PBH}+\frac14N_\mathrm{reh}(1-3w_\mathrm{reh})&-\frac12\ln\frac{H(N_\mathrm{PBH})}{H(N=0)}\nonumber\\
&=N_\mathrm{tot}-\Delta n_\mathrm{PBH},
\end{align}
where the first factor
\begin{align}
N_\mathrm{tot}&=n_\mathrm{CMB}+n_\mathrm{reh};\\
&=\ln\frac{T_\gamma}{H_0}-\ln\frac{k_\mathrm{CMB}}{a_0H_0}
+\ln\left(\frac{43}{11g_\mathrm{reh}}\right)^\frac13\left(\frac{\pi^2g_\mathrm{reh}}{90}\right)^\frac14\nonumber\\
&+\frac14\ln\left(\frac{\pi^2}{2}rA_s\right)\approx65+\frac14\ln\left(rA_s\right)
\end{align}
is constrained from the local large-scale physics at CMB scale independent of small-scale physics from reheating history, and the second factor 
\begin{align}
\Delta n_\mathrm{PBH}&\equiv n_\mathrm{CMB}-n_\mathrm{PBH};\\
&=N_\mathrm{CMB}-N_\mathrm{PBH}-\frac12\ln\frac{H(N_\mathrm{CMB})}{H(N_\mathrm{PBH})};\\
&\equiv\Delta N_\mathrm{PBH}-\frac12\int^{N_\mathrm{CMB}}_{N_\mathrm{PBH}}\epsilon_H(N)\mathrm{d}N
\end{align}
depends only on the local shape of inflationary potential from the CMB scale to the exit of fluctuations that later collapse into PBHs. Therefore, the mass of monochromatic PBH is immune from reheating history for fixed $\Delta N_\mathrm{PBH}\equiv N_\mathrm{CMB}-N_\mathrm{PBH}$. See the Fig.\ref{fig:monochromatic} and the very nice explaination in the appendix of \cite{Pi:2017gih} where $w_\mathrm{reh}=0$ in particular. However, the situation becomes different for PBH with extended mass fraction, because the relative abundance of each monochromatic component of PBHs with different $\Delta N_\mathrm{PBH}$ can changed dramatically for different reheating histories. 

For monochromatic PBH, the mass fraction at formation is defined by
\begin{align}\label{eq:betaform}
\beta^\mathrm{form}(M)=\frac{\rho_\mathrm{PBH}^\mathrm{form}(M)}{\rho_\mathrm{PBH}^\mathrm{form}(M)+\rho_\mathrm{rad}^\mathrm{form}(M)},
\end{align}
which is evolved to the radiation-matter equality according to
\begin{align}
\beta^\mathrm{eq}(M)=\frac{(a_\mathrm{eq}^{-3}/a_\mathrm{form}^{-3})\rho_\mathrm{PBH}^\mathrm{form}}{(a_\mathrm{eq}^{-3}/a_\mathrm{form}^{-3})\rho_\mathrm{PBH}^\mathrm{form}+(a_\mathrm{eq}^{-4}/a_\mathrm{form}^{-4})\rho_\mathrm{rad}^\mathrm{form}}.
\end{align}
After replacing $\rho_\mathrm{rad}^\mathrm{form}$ by \eqref{eq:betaform}, one obtains
\begin{align}
\beta^\mathrm{eq}(M)&=\frac{1}{1+\left(\frac{1}{\beta^\mathrm{form}(M)}-1\right)\frac{a_\mathrm{form}}{a_\mathrm{eq}}};\label{eq:betaeq}\\
&\approx\frac{a_\mathrm{eq}}{a_\mathrm{form}}\beta^\mathrm{form}(M),\quad \beta^\mathrm{form}(M)\ll1,
\end{align}
where the scale factor $a_\mathrm{eq}=1/3300$ and 
\begin{align}
a_\mathrm{form}=\frac{k(N_\mathrm{PBH})}{H(N_\mathrm{PBH})}\mathrm{e}^{2(n_\mathrm{PBH}+n_\mathrm{reh})}.
\end{align}
We will use the exact evaluation \eqref{eq:betaeq} instead of the approximation from small $\beta^\mathrm{form}(M)$ limit. The current observations constrain the abundance of PBH with respect to the quantity
\begin{align}
f=\frac{\Omega_\mathrm{PBH}^\mathrm{eq}}{\Omega_\mathrm{DM}^\mathrm{eq}}
\equiv\frac{\beta^\mathrm{eq}}{\Omega_\mathrm{DM}^\mathrm{eq}}
\approx\frac{\beta^\mathrm{eq}}{0.42}.
\end{align}

\subsection{The non-monochromatic case}\label{subsec:non-monochromatic}

For non-monochromatic PBH, the mass fraction at radiation-matter equality should be accumulated by
\begin{align}
\Omega_\mathrm{PBH}^\mathrm{eq}
=\int_{M_\mathrm{eva}}^{M_\mathrm{eq}}\frac{\mathrm{d}M}{M}\beta^\mathrm{eq}(M)
=\int_{N_\mathrm{eva}}^{N_\mathrm{eq}}\frac{\mathrm{d}M}{\mathrm{d}N}\frac{\mathrm{d}N}{M}\beta^\mathrm{eq}(M(N)),
\end{align}
where $M_\mathrm{eva}$ is the horizon mass of PBH at formation that exactly evaporates away totally at radiation-matter equality, and $M_\mathrm{eq}$ is the horizon mass of PBH formed at radiation-matter equality. Since the mass of PBH is a function of the e-folding number $N$ when the collapsed primordial fluctuations exit the Hubble horizon, one can find the e-folding number $N_\mathrm{eva}$ of exited primordial fluctuations that collapse into PBH of mass $M_\mathrm{eva}$ via
\begin{align}
M_\mathrm{PBH}^\mathrm{form}(N_\mathrm{eva})=\left(\frac{6\times10^4\,\mathrm{yrs}}{2.1\times10^{67}\,\mathrm{yrs}}\right)^\frac13M_\odot,
\end{align}
and one can also find the e-folding number $N_\mathrm{eq}$ of exited primordial fluctuations that later collapse into PBHs of mass $M_\mathrm{eq}$ via
\begin{align}
\frac{M_\mathrm{PBH}^\mathrm{form}(N_\mathrm{eq})}{M_\mathrm{PBH}^\mathrm{form}(N_\mathrm{eva})}
&\equiv\frac{M_\mathrm{eq}}{M_\mathrm{eva}}=\frac{H_\mathrm{eva}}{H_\mathrm{eq}}=\frac{a_\mathrm{eq}^2}{a_\mathrm{eva}^2};\\
&=\left(\frac{a_\mathrm{eq}}{\frac{k(N_\mathrm{eva})}{H(N_\mathrm{eva})}\mathrm{e}^{2(n_\mathrm{eva}+n_\mathrm{reh})}}\right)^2
\end{align}
with
\begin{align}
n_\mathrm{eva}\equiv N_\mathrm{eva}-\frac12\ln\frac{H(N_\mathrm{eva})}{H(N=0)}.
\end{align}
The abundance of PBH is then
\begin{align}
f=\frac{\Omega_\mathrm{PBH}^\mathrm{eq}}{\Omega_\mathrm{DM}^\mathrm{eq}}
\approx\frac{\Omega_\mathrm{PBH}^\mathrm{eq}}{0.42},
\end{align}
which cannot be simply compared with the current constraints on PBH that need to be modified for extended mass fraction \cite{Carr:2017jsz}.

\begin{figure}
\centering
\includegraphics[width=0.49\textwidth]{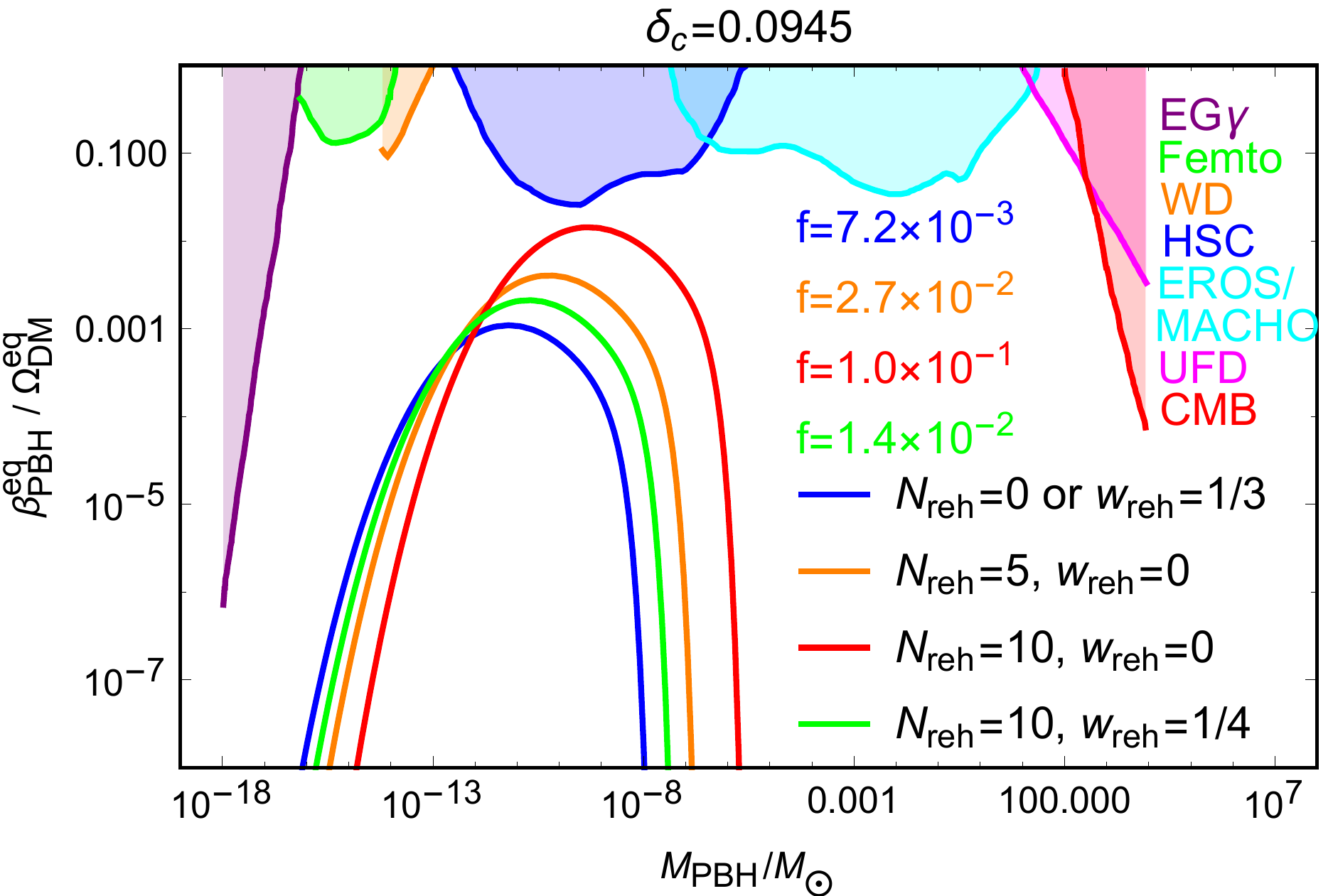}\\
\caption{The mass fraction of PBH in DM $\beta^\mathrm{eq}/\Omega_\mathrm{DM}^\mathrm{eq}$ at radiation-matter equality for given formation threshold $\delta_c=0.0945$. The corresponding abundance of PBH in DM $f=\Omega_\mathrm{PBH}^\mathrm{eq}/\Omega_\mathrm{DM}^\mathrm{eq}$ varies by many orders of magnitude for different reheating histories. The observational constraints are taken from the extragalactic photon (EG$\gamma$ \cite{Carr:2009jm}), femtolensing of gamma-ray burst (Femto \cite{Barnacka:2012bm}), white dwarf explosions (WD \cite{Graham:2015apa}), microlensing from Subaru Hyper Suprime-Cam (HSC \cite{Niikura:2017zjd}), EROS \cite{Tisserand:2006zx} and MACHO \cite{Allsman:2000kg}, ultra-faint dwarfs (UFD \cite{Brandt:2016aco}) and CMB \cite{Ali-Haimoud:2016mbv} (See also \cite{Poulin:2017bwe} for the CMB constraints on realistic disk-accretions onto PBHs).}\label{fig:PBH}
\end{figure}

\begin{figure}
\includegraphics[width=0.49\textwidth]{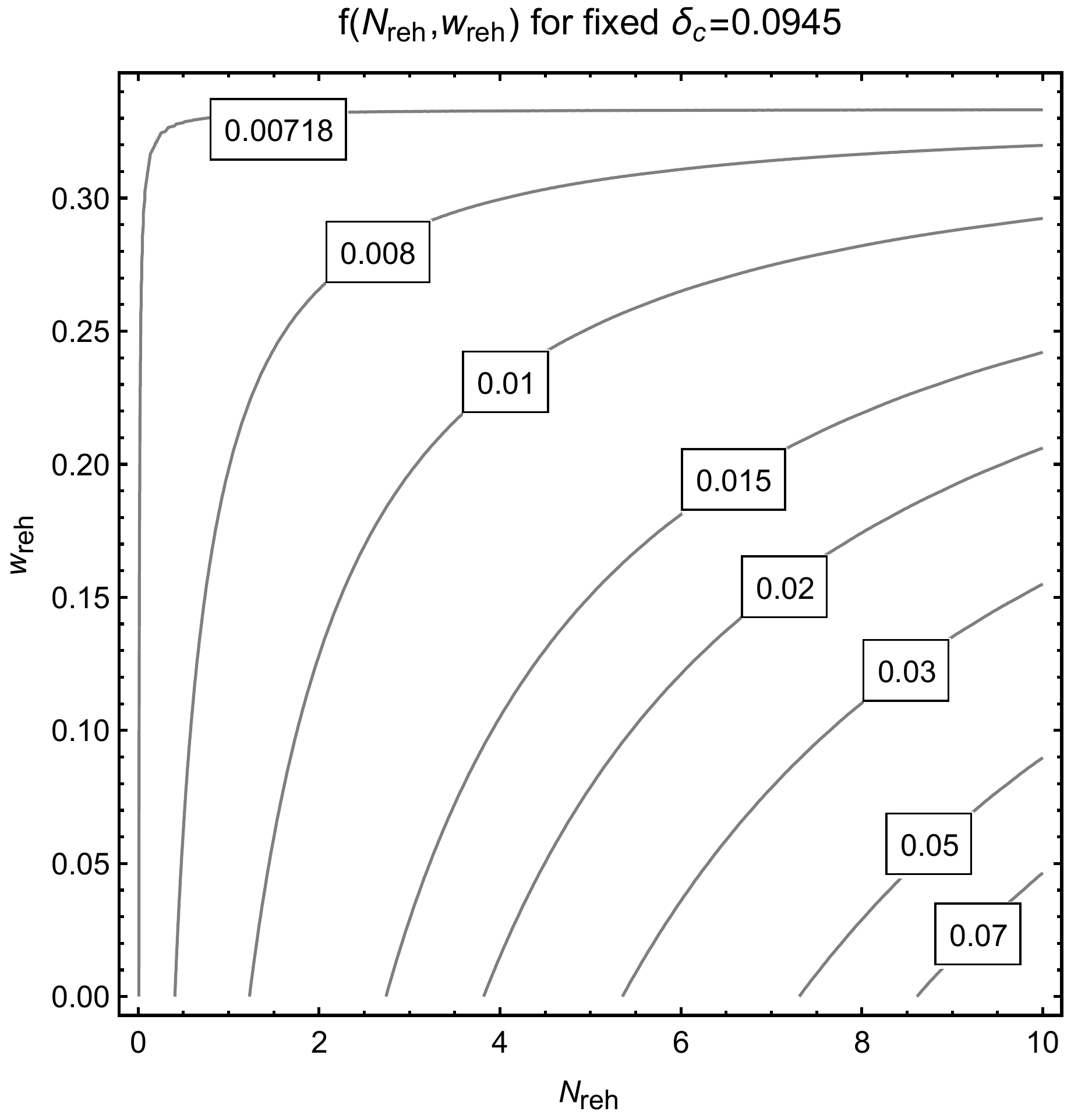}\\
\caption{The contour plot of PBH abundance $f=\Omega_\mathrm{PBH}^\mathrm{eq}/\Omega_\mathrm{DM}^\mathrm{eq}$ for given formation threshold $\delta_c=0.0945$ with respect to different reheating histories characterized by $N_\mathrm{reh}$ and $w_\mathrm{reh}$.}\label{fig:contour}
\end{figure}

At last, we present our final results in Fig.\ref{fig:PBH} and Fig.\ref{fig:contour}. The mass fraction of PBH in DM $\beta^\mathrm{eq}/\Omega_\mathrm{DM}^\mathrm{eq}$ at radiation-matter equality is presented in Fig.\ref{fig:PBH}, whose peak amplitude and position can vary by many orders of magnitude for different reheating histories with fixed formation threshold $\delta_c=0.0945$, and the corresponding abundance of PBH in DM $f=\Omega_\mathrm{PBH}^\mathrm{eq}/\Omega_\mathrm{DM}^\mathrm{eq}$ also varies by many orders of magnitude for different reheating histories. More specifically, both the mass fraction (peak amplitude and position) and corresponding abundance get increased for larger $N_\mathrm{reh}$ and fixed $w_\mathrm{reh}$, while get suppressed for larger $w_\mathrm{reh}$ and fixed $N_\mathrm{reh}$.
We further present in Fig.\ref{fig:contour} the PBH abundance $f=\Omega_\mathrm{PBH}^\mathrm{eq}/\Omega_\mathrm{DM}^\mathrm{eq}$ for given formation threshold $\delta_c=0.0945$,  which apparently varies by many orders of magnitude with respect to different reheating histories characterized by $N_\mathrm{reh}$ and $w_\mathrm{reh}$. Here the parameter ranges of reheating history are chosen as $0\leq N_\mathrm{reh}\leq10$ and $0\leq w_\mathrm{reh}\leq1/3$ for our specific potential employed in this work \footnote{We thank an anonymous referee for pointing this to us.}. Larger values for $w_\mathrm{reh}$ require potentials dominated by higher-dimensional operators like $\phi^6$.
It is worth noting that the precise values of formation threshold are not important for our purpose to show the sensitivity from different reheating histories on the abundance of PBH in DM.

\section{Conclusions} \label{sec:conclusion}

Recently there is a growing interest in the scenario of PBH as DM candidate. However, the predictions from PBH usually vary by orders of magnitude due to some exponential sensitivities on the model parameters, including generation models, formation models, accretion models and merger models. In this paper, we add another layer of uncertainty on the scenario of PBH as DM, which is the sensitivity from different reheating histories. Although, the abundance of PBH is independent of reheating history for monochromatic case, the abundance of PBH with extended mass fraction can vary many orders of magnitude for different reheating histories. We illustrate this reheating sensitivity in a simplest model of single-field inflation with a quasi-inflection point, which we believe is also manifest itself in other generation models of PBH, for example, double inflation model \cite{Kawasaki:2016pql,Inomata:2017okj,Inomata:2017vxo}, double inflection-point model \cite{Gao:2018pvq}, to name just a few. It is worth noting that the reheating sensitivity shown in this paper is different from those interesting discussions where PBHs are formed during slow reheating after inflation \cite{Carr:2018nkm} or during matter-dominated epoch after inflation but before reheating \cite{Carr:2017edp,Harada:2017fjm,Kohri:2018qtx}.

\begin{acknowledgements}
We thank Juan Garc\'{i}a-Bellido and Ester Ruiz Morales for the helpful correspondence and Lang Liu, Xue-Wen Liu, Shi Pi, Wu-Tao Xu for the helpful discussions. This work is supported by the National Natural Science Foundation of China Grants No.11690022, No.11375247, No.11435006, and No.11647601, and by the Strategic Priority Research Program of CAS Grant No.XDB23030100 and by the Key Research Program of Frontier Sciences of CAS. 
\end{acknowledgements}

\bibliographystyle{utphys}
\bibliography{ref}

\end{document}